\documentclass[12pt,a4paper]{article}
\usepackage{amsmath}
\usepackage{amsfonts}
\usepackage{amssymb}
\usepackage{makeidx}
\usepackage[pdftex]{graphicx}
\usepackage{textcomp}

\begin{document}

\title{\textbf{Entropy generation and jet engine optimization}}
\author{Umberto Lucia\\I.T.I.S. A. Volta\\Spalto Marengo 42, 15121 Alessandria, Italy} 
\date{}
\maketitle
 
\begin{abstract}
In 2009, it was shown that, with an original approach to
hydrodynamic cavitation, a phenomenological model was realized in order to compute some of the physical parameters needed for the design of the most common technological applications (turbo-machinery, etc.) with an economical saving in planning because this analysis could allow engineers to reduce the experimental tests and the consequent costs in the design process. Here the same approach has been used to obtain range of some physical quantity for jet engine optimization.
\end{abstract}

\section{Introduction}
The notions of entropy and its generation, both in equilibrium and in non-equilibrium processes, is the basis of the modern thermodynamics and statistical physics and engineering [\ref{bruers}]. It has been proved that entropy is a quantity which allows us to describe the progress of non-equilibrium dissipative process. The maximum entropy production principle, MEPP, has been introduced and used by several scientists and engineers throughout the 20th century when they dealt with general theoretical issues of technical and applied thermodynamics. By this principle, a non-equilibrium system develops following the thermodynamic
path which maximises its entropy production under the present constraints [\ref{lucia1}]. The second law of thermodynamics states
that for an arbitrary adiabatic process entropy of the final state is larger than or equal to entropy of the initial state; in terms of
the entropy generation, it means that the entropy production tends to a maximum. The MEPP may be viewed as the
natural generalisation of the Clausius-Boltzmann-Gibbs formulation of the second law [\ref{mart}]: ‘‘The relationship between the
minimum entropy production principle and MEPP is not simple; in fact, these variation principles are absolutely different:
although the extremum of one and the same function, the entropy production, is sought, these principles include different
constraints and different variable parameters. As a consequence, these principles should not be mutually opposed since they
are applicable to different stages of the evolution of a non-equilibrium system’’ [\ref{mart}]. 

It has been proved [\ref{bruers},\ref{8}-\ref{12}] that the MEPP can be considered as a universal principle governing the evolution of non-equilibrium dissipative systems [\ref{8}-\ref{12}]. In 2007
the principle of maximum entropy generation, elsewhere called maximum entropy variation due to irreversibility or maximum
irreversible entropy Sirr , was proved for open systems [\ref{lucia1},\ref{8}-\ref{12}], and in 2010 its statistical expression [\ref{lucia2}] has been introduced,
for a general open system, too. The problem of irreversibility is difficult and part of this difficulty is due to dealing with the
statistical mechanics of a large number of particles [11], but a global approach has been introduced in 2008 [10].

Applications of the maximum entropy generation principle have been done in
biophysics, in molecular physics, in technical thermodynamics, in fluid flow analysis [\ref{lucia3},\ref{lucia4}].

In ref. [\ref{riggins}] provides the development of the expressions for jet engine specific thrust and specific impulse in terms of fundamental
thermodynamic quantities, including the irreversibility occurring within the
actual engine flow-field. The continuum of performance provides a powerful tool for understanding, optimizing, and assessing engine types, regimes, and
performance issues from a thrust-based performance perspective. It also provides the natural base-line for measuring the impact of irreversibility on engine performance. It pointed out a very simple fluids problem in which
conventional flow availability as suggested in numerous references fails to yield the optimized configuration. The conclusion is drawn that conventional flow availability analysis needs to be revisited, at least in terms of the functionality or purpose of the vehicle and how an availability analysis should be applied to the optimization of the vehicle. Moreover, Riggins underlined that any proposed candidate thermodynamic currency for vehicle optimization must be robust in terms of also achieving optimization for
isolated sub-systems with given constraints. He reviewed the meaning and
evaluation of lost thrust work from fundamental thermodynamic principals and then argues for an extension of the concept to the broader principle
of the minimization of lost work between actual and reversible devices in system optimization efforts. This result was based on the successful
linkage established between engine performance and lost thermodynamic work for engine-only applications [\ref{riggins}]. One of the main components of a multirole aircraft, determining it's flight performance characteristics, is the propulsion system. At selection of aerodynamic arrangement parameters and jet engine working process parameters there are complexities stipulated by an inconsistency of sub- and super-sonic flight regimes. Direct method of aerodynamic calculation - from geometry to the flight performance and maneuvering characteristics is characterized by search of large number of versions and requires large costs of time. The essential simplification of the problem is provided by use of return way of aerodynamic calculation - from the point flight-performance (PFP) to required parameters of arrangement and engine. Therefore the method of simultaneous aerodynamic arrangement parameters (size, geometry) and engine parameters (size, bypass ratio etc) selection under design requirements to the aircraft is actual [\ref{c2}].

In this paper we will develop a thermodynamic optimization of jet engine starting from the the maximum entropy generation principle.

\section{Entropy and irreversible entropy}
The open thermodynamic system has been analytically introduced in terms of advanced analysis in [\ref{12}]. Here its phenomenological description is proposed as follows:
\newtheorem{definition}{Definition}
Let us consider an open continuum or discrete $N$ particles system. Every $i-$th element of this system is located by a position vector $\mathbf{x}_{i}\in \mathbb{R}^{3}$, it has a velocity $\Dot{\mathbf{x}}_{i}\in \mathbb{R}^{3}$, a mass $m_{i}\in \mathbb{R}$ and a momentum $\mathbf{p}_{i}=m_{i}\mathbf{\Dot{x}}_{i}$, with $i\in [1,N]$ and $\mathbf{p}\in \mathbb{R}^{3}$ [\ref{lucia1}]. The masses $m_{i}$ must satisfy the condition:
	\begin{equation}
		\sum_{i=1}^{N}m_{i}=m
	\end{equation} 
where $m$ is the total mass which must be a conserved quantity so that it follows:
	\begin{equation}
		\Dot{\rho}+\rho \nabla\cdot\Dot{\mathbf{x}}_{B}=0
	\end{equation} 
where $\rho =dm/dV$ is the total mass density, with $V$ total volume of the system and $\Dot{\mathbf{x}}_{B}\in \mathbb{R}^{3}$, defined as $\Dot{\mathbf{x}}_{B}=\sum_{i=1}^{N}\mathbf{p}_{i}/m$, velocity of the centre of mass. The  mass density must satisfy the following conservation law [\ref{lucia1}]:
	\begin{equation}
		\Dot{\rho}_{i}+\rho_{i} \nabla\cdot\Dot{\mathbf{x}}_{i}=\rho \Xi
	\end{equation} 
where $\rho_{i}$ is the density of the $i-$th elementary volume $V_{i}$, with $\sum_{i=1}^{N}V_{i}=V$, and $\Xi$ is the source, generated by matter transfer, chemical reactions and thermodynamic transformations.

We follow a global approach as usual done in engineering thermodynamics: the usual analysis is based on the variation of the entropy (in engineering thermodynamics it is well defined by means of global quantities) and considers two components for the entropy: one related to external exchange $d_{ex}S=\delta Q/T$, with $Q$ heat exchanged and $T$ temperature, and the other related to the internal origin $d_{in}S=-X d\alpha$, with $X$ non-conservative forces and $alpha$ extensive thermodynamic quantities. Then $dS=d_{ex}S+d_{in}S$ and the equation of entropy balance for the system results [\ref{lucia1}]:
\begin{equation}
\int_{V}\rho \frac{ds}{dt}dV+\int_{V}\nabla\cdot\mathbf{J}_{S}dV=\frac{dS}{dt}
\end{equation}
where $s$ is the entropy density and $\mathbf{J}_{S}=\left(\mathbf{Q}/T\right)+\sum_{i}\rho_{i}s_{i}\left(\Dot{\mathbf{x}}_{i}-\Dot{\mathbf{x}}_{B}\right)$ is the entropy flux, with $\textbf{Q}$ heat flux, $\rho$ density, $\textbf{x}$ position and the suffix $B$ means centre of mass. It was proved that, on the analysis of the irreversible systems, the entropy related to the internal origin can be expressed by the irreversible entropy [\ref{lucia1},\ref{12}]. This result means that $\int d_{in}S=\Delta S_{irr}=T_{a}W_{lost}$, where the irreversible entropy can be defined following a global approach as:
\begin{equation}
\Delta S_{irr}=\frac{Q_{r}}{T_{a}}\left(1- \frac{T_{a}}{T_{r}}\right)+\frac{\Delta H}{T_{a}}-\Delta_{ex} S+\frac{\Delta E_{k}+\Delta E_{g}-W}{T_{a}}
\label{irr}
\end{equation}
where $Q_{r}$ is the heat source, $T_{r}$ its temperature, $T_{a}$ the ambient temperature (in general reference temperature, i.e. the least reservoir temperature), $H$ is the enthalpy, $\Delta_{ex}S= S_{ing}-S_{outg}=\int d_{ex}S$, with $S_{ing}$ entropy which enters into the system and $S_{outg}$ entropy which flows out of the system, $E_{k}$ the kinetic energy, $E_{g}$ the gravitational one, $W$ the work done and $W_{lost}$ work lost for irreversibility. 

In equilibrium and non-equilibrium thermodynamics the concept of thermostat is very important: it allows us to represent the mechanism which enables the system to reach a non-equilibrium stationary state in the presence of an imposed external force. In a mechanical statistical approach to the non-equilibrium stationary states for a system subject to non-conservative external forces, it has been proved that the statistical and global approach converge to the same results [\ref{lucia3}]. So the global approach allows us to state that relations as $\int\left(dU + p dV\right)/T=$exact are possible also in stationary non-equilibrium states; indeed, we can write that $\Delta S=\Delta_{ex}S+\Delta S_{irr}$, but $\delta Q=dU + p dV$ in every real global system, so that:
\begin{equation}
\begin{split}
\int\frac{\left(dU + p dV\right)}{T}=\int\frac{\delta Q}{T}=\Delta_{ex} S &=\frac{Q_{r}}{T_{a}}\left(1- \frac{T_{a}}{T_{r}}\right)+\frac{\Delta H}{T_{a}}-\frac{W_{lost}}{T_{a}}+\\ &+\frac{\Delta E_{k}+\Delta E_{g}-W}{T_{a}}
\end{split}
\label{entropyex}
\end{equation}

Moreover, for an open thermodynamic system, it has been proved the principle of maximum irreversible entropy as a macroscopic and global approach; it states that the condition of stability for the open system is that its irreversible entropy variation $\Delta S_{irr}$ reaches its maximum:
\begin{equation}
\delta\left(\Delta S_{irr}\right)\geq 0
\label{maxent}
\end{equation}

\section{Jet engine optimization}
Let us consider a jet engine. Its specific enthalpy variation can be obtained as [\ref{kirillin}]:
\begin{equation}
h_2 - h_1 = \frac{1}{2}(w_2^2 - w_1^2)
\end{equation}
where $h$ is the specific enthalpy and $w$ the velocity of fluid ejected. Using the definition of specific enthalpy $h_2 - h_1 = T\,(s_2 - s_1)+v\,(p_2-p_1)$ it follows that:
\begin{equation}
T\,(s_2 - s_1)+v\,(p_2-p_1) = \frac{1}{2}(w_2^2 - w_1^2)
\end{equation}
where $T$ is the temperature, considered constant as a consequence of the Gouy-Stodola theorem [\ref{kirillin}], $s$ is the specific entropy, $p$ the pressure and $w$ the velocity of the fluid ejected. Consequently, the efficiency $\eta$, defined in Ref. [\ref{kirillin}], results:
\begin{equation}
\eta = \dfrac{1}{1+\dfrac{c_p\,T_1}{\dfrac{(w_2^2-w_1^2)}{2}}}=\dfrac{1}{1+\dfrac{c_p\,T_1}{T\,(s_2 - s_1)+v\,(p_2-p_1)}}
\end{equation}
with $c_p$ specific heat at constant pressure. From this last relation the specific entropy variation can be obtained as:
\begin{equation} \label{entropy}
s_2 - s_1 = \frac{1}{T}\,\Bigg[\frac{c_p\,T_1}{\frac{1}{\eta}-1}-v\,(p_2-p_1)\Bigg]
\end{equation}
but, considering that [\ref{kirillin}]:
\begin{equation}
\eta = 1 - \dfrac{1}{\beta^{\frac{k-1}{k}}}
\end{equation}
with $\beta=p_2 /p_1$ compression ratio, the relazion (\ref{entropy}) becomes:
\begin{equation}
s_2 - s_1 = \frac{1}{T}\,\Bigg[\frac{c_p\,T_1}{\frac{1}{1-\beta^{\frac{k-1}{k}}}-1}-v\,(p_2-p_1)\Bigg]=\frac{1}{T}\,\big[c_p\,T_1\,\big(\beta^{\frac{k-1}{k}}-1\big)-v\,(p_2 - p_1)\big]
\end{equation}
Now, considering the maximum entropy generation principle [\ref{8}-\ref{12}] it follows that:
\begin{equation}
c_p\,T_1\,\frac{k-1}{k}\,\frac{1}{\beta^{1/k}}-v\,p_1= 0
\end{equation}
from which the following condition can be obtained:
\begin{equation}
\beta = \Bigg(\frac{c_p\,T_1}{v\,p_2}\;\frac{k-1}{k}\Bigg)^{\frac{k}{k-1}}
\end{equation}

\section{Conclusions}
Reducing mass, fuel consumption and development cost while increasing performance are the traditional drivers of aircraft engine design. Significantly reduced environmental impact is an additional, and increasingly important, design objective. Novel engine concepts and architectures are being explored and investigated to meet these challenges.
As component manufacturers supply jet engine parts to original equipment manufacturers, it is desirable to design components using a whole jet engine approach in order to optimize component design for system-level performance. There are, however, several issues that hinder this approach in current
practice. Component manufacturers have a need for models that effectively integrate the product definition with the analysis activities during early product development activities. Aircraft design is traditionally strong in modeling and simulation but model integration and minimization of redundant information is still relatively weak due to the advanced, but domain-specific and methods used. Moreover, component manufacturers are often forced to work with engine system and component simulation
models that have different levels of fidelity and are not integrated. There is therefore a need to strengthen the capability to model and understand the engine component behavior in the entire jet engine system context. The
properties of engine components must be optimized to satisfy targets that are set on engine system level, rather than on properties derived at component-system boundaries [\ref{c1}].

The efficiency of a jet engine is strongly dependent on the pressure drop through the turbine and nozzle. To achieve the largest possible drop, the engine operates at the highest possible compression ratio. Higher compression ratios imply higher compressor outlet temperatures and thus higher flame temperatures. The tolerable temperature limit is set by the turbine blades— usually the first stage. Modern turbine blades are single metal crystals with hollow interiors. Cooler air from the compressor is blown through the hollow interior of the blades. In a modern engine the turbine inlet temperature will typically be around 1700\textcelsius, higher than the melting temperature of the blade material (around 1600\textcelsius). Still higher temperature operation will require not only better materials but also some means of eliminating the oxides of nitrogen that form at such high combustion temperatures. Today these problems are much better handled but temperature still limits airspeeds in supersonic flight. At the very highest speeds the compression of the intake air raises the temperature to the point that the compressor blades will melt. At lower speeds better materials have increased the critical temperature and automatic fuel management controls have made it nearly impossible to overheat the engine. Turbojets do not throttle efficiently. To operate well at all the compressor blades must turn at not less than 50 to 70\% of the design maximum. At low throttle settings a great deal of power is wasted compressing a large fraction of the full-throttle airflow, only to expand it back again with relatively little temperature gain from the combustion chamber. Poor efficiency at low throttle settings helps to explain why turbines aren't used in cars—the engine would be burning a huge quantity of fuel even while sitting at a red light. In aircraft every bit of efficiency in running the compressor is needed. One common design technique is to use more than one turbine to drive the compressor stages at various speeds.

In 2009 it was shown that, with an original approach to
hydrodynamic cavitation [\ref{lucia4}], a phenomenological model could
be realized in order to compute some of the physical parameters needed for the design of the most common technological applications (turbo-machinery, etc.) with an economical saving in planning because this analysis could
allow engineers to reduce the experimental tests and the consequent costs in the design process [\ref{lucia4}].

Following the same approach a thermodynamic model has been proposed to develop an engineering and technical physics analysis of jet engine in order to obtain range of some physical quantity for their optimization.

\bibliographystyle{unsrt}
%\bibliography{bibfile}

%\newpage

\end{document}